\pdfoutput=1
\documentclass{IEEEtran}

\usepackage[utf8]{inputenc} 
\usepackage[T1]{fontenc}    
\usepackage{hyperref}       
\usepackage{url}            
\usepackage{booktabs}       
\usepackage{amsfonts}       
\usepackage{amsmath}
\usepackage{nicefrac}       
\usepackage{microtype}      
\usepackage{float}
\usepackage[pdftex]{graphicx}
\usepackage{color}
\usepackage{amssymb}
\usepackage{bm}
\usepackage{multirow}
\usepackage{svg}
\usepackage{graphicx}
\usepackage[shortcuts]{extdash}
\usepackage{subfig}

\usepackage[font=small,labelfont=bf]{caption}

\usepackage{algpseudocode}

\usepackage[textsize=tiny,textwidth=1.2cm]{todonotes}

\hypersetup{hidelinks}

\def\tablename{Table}

\begin{document}

\title{Detection of prostate cancer in whole-slide images through end-to-end training with image-level labels}

\author{Hans Pinckaers*, Wouter Bulten, Jeroen van der Laak, Geert Litjens
\thanks{Hans Pinckaers, Wouter Bulten, Jeroen van der Laak and Geert Litjens are with the Computational Pathology Group, Department of Pathology, Radboud Institute for Health Sciences, Radboud University Medical Center, Nijmegen, The Netherlands; Additionally, Jeroen van der Laak is with Center for Medical Image Science and Visualization, Linköping University, Linköping, Sweden. {\it Asterisk indicates corresponding author. } e-mail: hans.pinckaers@radboudumc.nl.}
}

\maketitle

\begin{abstract}
Prostate cancer is the most prevalent cancer among men in Western countries, with 1.1 million new diagnoses every year. The gold standard for the diagnosis of prostate cancer is a pathologists' evaluation of prostate tissue.

To potentially assist pathologists deep\=/learning\=/based cancer detection systems have been developed. Many of the state-of-the-art models are patch\=/based convolutional neural networks, as the use of entire scanned slides is hampered by memory limitations on accelerator cards. Patch-based systems typically require detailed, pixel-level annotations for effective training. However, such annotations are seldom readily available, in contrast to the clinical reports of pathologists, which contain slide-level labels. As such, developing algorithms which do not require manual pixel-wise annotations, but can learn using only the clinical report would be a significant advancement for the field.

In this paper, we propose to use a streaming implementation of convolutional layers, to train a modern CNN (ResNet\=/34) with 21 million parameters end-to-end on 4712 prostate biopsies. The method enables the use of entire biopsy images at high-resolution directly by reducing the GPU memory requirements by 2.4 TB. We show that modern CNNs, trained using our streaming approach, can extract meaningful features from high-resolution images without additional heuristics, reaching similar performance as state-of-the-art patch-based and multiple-instance learning methods. By circumventing the need for manual annotations, this approach can function as a blueprint for other tasks in histopathological diagnosis.

The source reproduce the streaming models is available at \url{https://github.com/DIAGNijmegen/pathology-streaming-pipeline}.

\end{abstract}

\section{Introduction}

The current state-of-the-art in computer vision for image classification tasks are convolutional neural networks (CNNs). Commonly, convolutional neural networks are developed with low\=/resolution labeled images, for example 0.001 megapixels for CIFAR-10\cite{Krizhevsky2009}, and 0.09-0.26 megapixels for ImageNet\cite{Russakovsky2015}. These images are evaluated by the network and the parameters are optimized with stochastic gradient descent by backpropagating the classification error. Neural networks learn to extract relevant features from their input. To effectively learn relevant features, optimizing these networks requires relatively large datasets.\cite{Sun2017RevisitingUE}

In histopathology, due to the gigapixel size of scanned samples, generally referred to as whole-slide images (WSIs), the memory limitation of current accelerator cards prohibits training on the entire image, in contrast to most of the natural images used in general computer vision tasks. As such, most networks are trained on tiny patches from the whole\=/slide image. Acquiring labels for these patches can be expensive. They are generally based on detailed outlines of the classes (e.g., tumor regions) by an experienced pathologist. This outlining is not done in clinical practice, and is a tedious and time-consuming task. This limits the dataset size for training models. Also, we will need to create these annotations for every individual task.

However, if we could circumvent labeling on a patch level, clinically evaluated biopsies could be cheaply labeled using their clinical reports. These reports contain all relevant information for clinical decisions, and are thus of large value for machine learning algorithms.

In this paper we will focus on prostate cancer detection. The diagnosis of prostate cancer---the most prevalent cancer for men in Western countries---is established by detection on histopathological slides by a pathologist. The microscopy slides containing cross-sections of biopsies can exhibit morphological changes to prostate glandular structures. In low-grade tumors, the epithelial cells still form glandular structures; however, in the case of high-grade tumors, the glandular structures are eventually lost\cite{Fine2012}.

In the presence of cancer, the percentage of cancerous tissue in a prostate biopsy can be as low as 1\%, the evaluation of the biopsies can be tedious and error-prone, causing disagreement in the detection of prostate cancer, as in the grading using the Gleason scheme\cite{Ozkan2016}.

Besides substantial inter-observer and intra-observer variability, diagnosing prostate cancer is additionally challenging due to increasing numbers of biopsies as a result of the introduction of prostate-specific antigen (PSA) testing\cite{10.1093/jnci/djp278}. This number is likely to increase further due to the aging population. In the light of a shortage of pathologists\cite{Wilson2018}, automated methods could alleviate workload.

To reduce potential errors and workload, recent work\cite{Bulten2020, Campanella2019, Nagpal2019, Litjens2016, Arvaniti2018, Lucas2019, Strom2020}, has shown the potential to automatically detect prostate cancer in biopsies. These studies either use expensive, pixel-level annotations or train CNNs with slide-level labels only, using a patch-based approach. 

One popular strategy is based on multiple-instance-learning (MIL)\cite{Courtiol2018,Ilse2018,Amores2013}. In this approach, the whole-slide image (WSI) is subdivided into a grid of patches. The MIL assumption states that in a cancerous slide ('positive bag'), at least one patch will contain tumorous tissue, whereas negative slides have no patches containing tumour. Under this assumption, a CNN is trained on a patch-level to find the most tumorous patch. 

However, this approach has several disadvantages.\cite{VanderLaak2019} First, this method only works for tasks where the label can be predicted from one individual patch and a single adversarial patch can result in a false positive detection. Second, it is essentially a patch-based approach, therefore, the size of the patch constrains the field-of-view of the network.

In this paper, we propose a novel method, using streaming\cite{Pinckaers2019}, to train a modern CNN (ResNet-34) with 21 million parameters end-to-end to detect prostate cancer in whole-slide images of biopsies. This method does not suffer from the same disadvantages as the aforementioned approaches based on MIL: it can use the entire content of the whole-slide image for its prediction and the field-of-view is not limited to an arbitrary patch-size. We compare our approach against the methods by Campanella et al.\cite{Campanella2019} and Bulten et al.\cite{Bulten2020}.

The streaming implementation allows us to train a convolutional neural network directly on entire biopsy images at high-resolution (268 megapixels) using only slide-level labels. We show that a state-of-the-art CNN can extract meaningful features from high-resolution images using labels from pathology reports without additional heuristics or post-processing. Subsequently, we show that transfer learning from ImageNet performs well for images that are 5000x bigger than the original images used for training (224x224).

\begin{figure*}[t]
    \centering
    {\includegraphics[width=\textwidth]{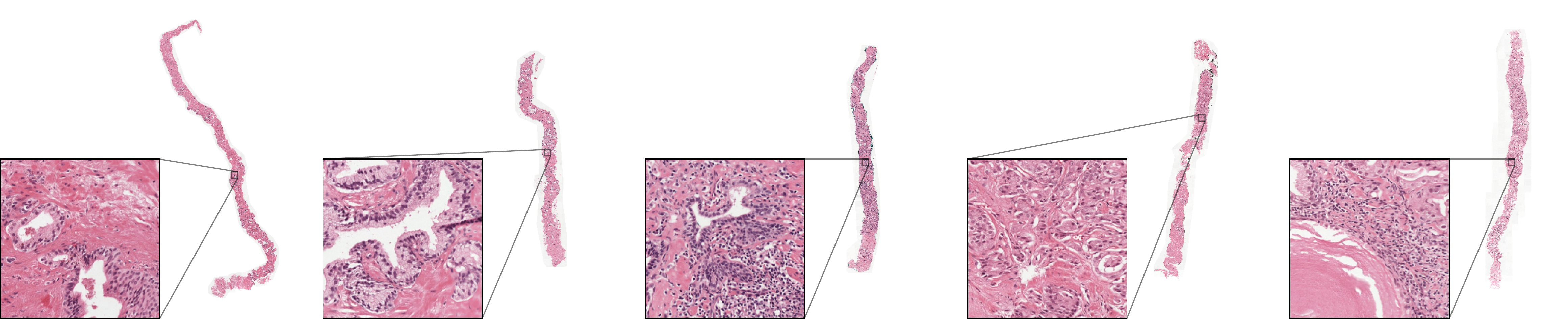}}%
    \caption{Example biopsies of our dataset. The left two biopsies are benign, the others cancerous. The zoomed regions show tumorous tissue when present. They are rendered in same resolution ($1.0 \mu m$/px, region is 328$\times$328 pixels) on which the models are trained. Best viewed digitally and at a high magnification.}
    \label{figure:streamingSGD}
\end{figure*}

\section{Related works}

For prostate cancer detection, previous works have used more traditional machine learning (i.e., feature-engineering) approaches\cite{Gertych2015, nguyen2017, Naik2007}. Recently, researchers transitioned to using deep-learning-based methods for the detection of cancer\cite{Campanella2019, Litjens2016}. Besides detection, research on prostate cancer grading has also been published\cite{Arvaniti2018, Lucas2019, Bulten2020}.

In this work, we train on labels for individual biopsies. Since in other work, the memory of the accelerator restricts the input size of the image, published methods are based on searching relevant patches of the original slide\cite{Ianni2020, Campanella2019, lu2020data, Li2019a}, or compressing the slide into a smaller latent space.\cite{Tellez2019}

We explicitly compare against the state-of-the-art method from Campanella \textit{et al.}\cite{Campanella2019}. As mentioned before, their multiple-instance-learning approach is based on the single most-informative patch, and thus leads to a small field-of-view for the network, and potential false positives because of a few adversarial patches. To circumvent some of these problems, Campanelle \textit{et al.}\cite{Campanella2019}, tried to increase the field-of-view to multiple patches using a recurrent neural networks with some improvement. Their system achieved an area-under-the-receiver-operating curve (AUC) of 0.986. the aggregation method increased the AUC to 0.991. To make the comparison fair, we trained an identical network architecture for both methods. However, when training end-to-end, the context of the whole image is automatically taken into account.

Campanella et al. showed that performance decreases when using smaller datasets, concluding that at least 10,000 biopsies are necessary for a good performance. Since they did not use data augmentation (probably because of the big dataset at hand), we investigated if we could reach similar performances with smaller dataset sizes using data augmentation.

Since the mentioned implementation of multiple-instance-learning only considers one patch, which may be less efficient, others\cite{lu2020data, Li2019a} improved the method by using multiple resolution patches and attention mechanisms. Li \textit{et al.} trained two models on low and high resolution patches, only patches that were predicted as suspicious by the lower resolution model were used to train the higher resolution model. Additionally, to calculate the attention mechanisms, all patches need to be kept in memory, limiting the size of the patches. Lu\textit{et al.} \cite{lu2020data} showed that, additionally to attention mechanisms, a frozen model pretrained on ImageNet decreases training time and improves data efficiency. We also use ImageNet weights, but by using the streaming-implementation of convolution, can unfreeze the model and train the whole network end-to-end. However, in both papers, no comparison to the original method of Campanella \textit{et al.} was performed. 

\section{Materials}

We used the same dataset as Bulten et al.\cite{Bulten2020}, we will briefly reiterate the collection of the dataset here. We built our dataset by retrospectively collecting biopsies and associated pathology reports of patients. Subsequently, we divided the patients between training, tuning, and test set. As standard practice, we optimized the model using the training set and assessed generalization using the tuning set during development. After development, we evaluated the model on the test set. The dataset, except for the test set, is publicly available as a Kaggle challenge at https://www.kaggle.com/c/prostate-cancer-grade-assessment.

\subsection{Data collection}
We retrieved pathologists reports of prostate biopsies for patients with a suspicion of prostate cancer, dated between Jan 1, 2012, and Dec 31, 2017, from digital patient records at the Radboud University Medical Center, excluding patients who underwent neoadjuvant or adjuvant therapy. The local ethics review board waived the need for informed consent (IRB approval 2016–2275). 

After anonymization, we performed a text search on the anonymized pathology reports to divide the biopsies into positive and negative cases. Afterward, we divided the patient reports randomly into training, tuning, and test set. By stratifying the biopsies on the primary Gleason score, we retrieved a comparable grade distribution in all sets. From the multiple cross-sections which were available per patient, we selected the standard hematoxylin-and-eosin-stained glass slide containing the most aggressive or prevalent part of malignant tissue for scanning.

We digitized the selected glass slides using a 3DHistech Pannoramic Flash II 250 (3DHistech, Hungary) scanner at a pixel resolution of $0.24 \mu m$. Since each slide could contain one to six unique biopsies, commonly with two consecutive sections of the biopsies per slide, trained non-experts coarsely outlined each biopsy, assigning each with either the reported Gleason score, or labeling negative, based on the individual biopsy descriptions in the pathology report.

We collected 1243 glass slides, containing 5759 biopsies sections. After division, the training set consisted of 4712 biopsies, the tuning set of 497 biopsies, and the test set of 550 biopsies.  We extracted the individual biopsies from the scanned slides at a pixel resolution of $0.96 \mu m$, visually approximately equivalent to 100x total magnification (i.e., 10x microscope objective with a standard 10x ocular lens). Subsequently, we trimmed the whitespace around the tissue using a tissue-segmentation neural network\cite{Bandi2019a}.

\subsection{Reference standard test set}

To determine a strong reference standard, three specialized pathologists reviewed the slides in three rounds. In the first round, each pathologist graded the biopsies independently. In the second round, each biopsy for which no consensus was reached in the first round, consensus was regraded by the pathologist whose score differed from the other two, with the help of the pathologist's first score and the two anonymous Gleason scores of the other pathologists. In the third round, the pathologists discussed the biopsies without consensus after round two. In total 15 biopsies were discarded by the panel as they could not be reliably graded, resulting in a total test set size of 535 biopsies. See\cite{Bulten2020} for a complete overview of the grading protocol.

\subsection{Smaller subsampled training set}

To test our method with smaller datasets, we sampled 250 (5\%) and 500 (10\%) biopsies from the training set. Half of the cases in the new sets were negatives. For the positive biopsies, we stratified on primary Gleason grade and sampled equal amounts of each. Thus, we kept the distribution of the positive biopsies equal over all the datasets. We used the 5\% (250 biopsies) and 10\% (500 biopsies) datasets for training. The tuning- and test-sets were equal to the ones used in the development of the model on the whole set. 

\begin{table}[h]
\footnotesize
\centering
\captionsetup{width=.75\linewidth,justification=centering}
\caption{Distribution of datasets used in the experiments, stratisfied on primary Gleason pattern.}
\label{tab:extset}
\begin{tabular}{llllll}
\toprule
\textbf{Dataset} & \textbf{Total} & \textbf{Negative} & \textbf{3} & \textbf{4} & \textbf{5} \\
\midrule
Training set & 4712 & 16\% & 32\% & 45\% & 7\% \\
Tuning set & 497 & 39\% & 23\% & 29\% & 9\% \\
\midrule
10\% set & 500 & 50\% & 17\% & 17\% & 17\% \\
5\% set & 250 & 51\% & 16\% & 16\% & 16\% \\
\midrule
Test set & 535 & 47\% & 25\% & 19\% & 9\% \\
\bottomrule
\end{tabular}
\end{table}

\section{Methods}

\subsection{End-to-end streaming model}

We trained a ResNet-34\cite{He2016} convolutional neural network. Since the individual biopsy images differ in size, we padded or center\-/cropped them to 16384$\times$16384 input. 99\% of our dataset biopsies fitted within this input size.

For regularization, we used extensive data augmentation. To make augmentation of these images feasible with reasonable memory usage and speed, we used the open-source library VIPS\cite{VIPS1996}. Elastic random transformation, color augmentation (hue, saturation, and brightness), random horizontal and vertical flipping, and rotations were applied. We normalized the images based on training dataset statistics.

For our streaming experiments, we initialized the network using ImageNet-trained weights. As an optimizer, we used standard SGD (learning rate of $2e-4$) with momentum (0.9) and a mini-batch size of 16 images. Because when using streaming, we do not have a full image on the GPU, we cannot use batch normalization, thus we froze the batch normalization mean and variance, using the transfer-learned ImageNet running mean and variance. We randomly oversampled negative cases to counter the imbalance in the dataset.\cite{Buda2018}

\begin{figure*}[h!]

      \centering
      {\includegraphics[width=\textwidth,height=150pt]{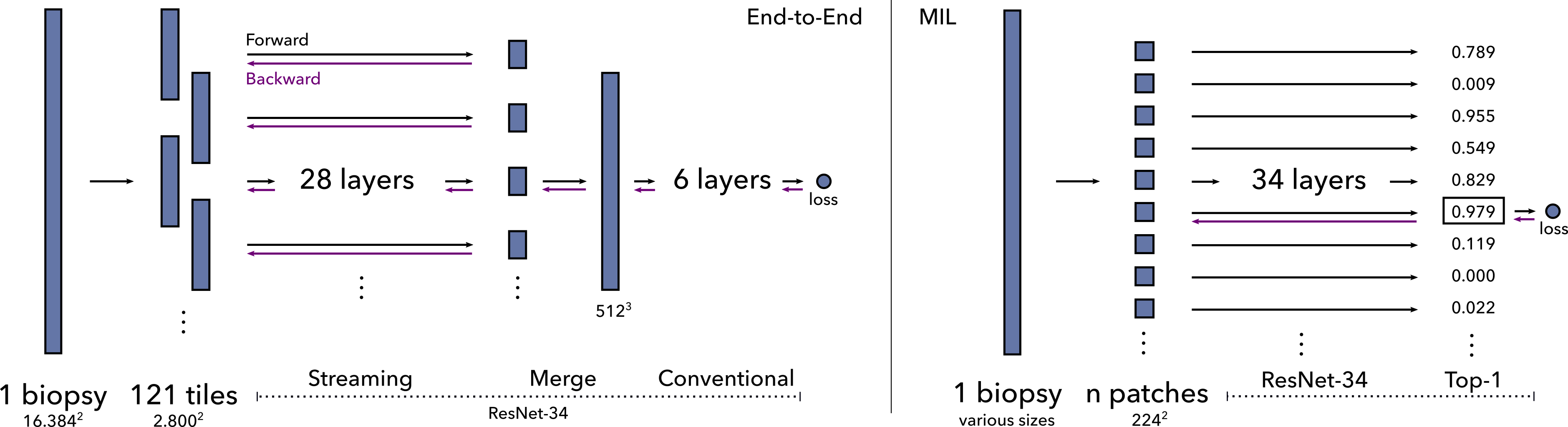}}%
      \caption{Using the streaming implementation of convolutional operations we can train a residual neural network end-to-end on whole-slide images (left). Streaming combines precise tiling with gradient checkpointing to reduce memory requirements. We can then receive gradient signal over the whole input image. Multiple-instance-learning (MIL) (right) divides the images into a grid of smaller patches with the assumption that an individual patch could determine the image-level label. The network is then only optimized using the patch with the highest probability.}
      \label{figure:streamingSGD}
\end{figure*}

\subsubsection{Training schedule}

In transfer learning, often the first layers are treated as a feature extraction algorithm. After the feature extraction part, the second part is trained for the specific task\cite{TanSKZYL18}. Since the domain of histopathology differs significantly from the natural images in ImageNet, we froze the first three (of the four) residual blocks of the network (the first 28 layers) as feature extractor, only training the last block for our task. This also has the benefit of training faster, since we do not need to calculate gradients for the first layers. After 25 epochs, all the networks were stabilized and stopped improving the tuning loss, showing slightly lower train losses.

From these epochs, we picked a checkpoint with a low tuning loss to resume fine-tuning the whole network, unfreezing the weights of the first three residual blocks. Due to the relatively small tuning set, the loss curve was less smooth than the training loss curve. To account for a sporadic checkpoint with a low loss, we calculated a moving average over five epochs. From these averages, we picked the window with the lowest loss, taking the middle checkpoint of the averaging window.

Starting from this checkpoint, we fine-tuned the whole network with a learning rate of $6e-5$. After approximately 50 epochs, all the networks stopped improving. For inference, we choose the checkpoints based on a moving average of five epochs with the lowest tuning set loss. We averaged the weights of these checkpoints to improve generalization\cite{Izmailov2018}.

The optimization and training procedure was fully conducted using the tuning set, the test was untouched during the development of the model.

\subsubsection{Streaming CNN}

To train a ResNet-34 with such high-resolution, we used our previously published method termed `streaming' as described in\cite{Pinckaers2019}. Streaming combines precise tiling with gradient checkpointing to reduce memory requirements.

In short, to replicate a forward pass of a high-resolution image, streaming involves the calculation of a feature map of a chosen layer somewhere mid-network, in a memory-efficient way. Due to downsampling, the feature map of this layer will be smaller and able to fit into GPU memory. We call this the \textit{reconstruction} of an intermediate feature map. This feature map will be identical, as would be the case if we had enough memory to calculate it from the original image. 

Since ResNet is a fully convolutional neural network, we accomplish the reconstruction by performing forward passes with tiles of the original image up until the layer of choice. We merge the outputs of each tile correctly to reconstruct the intermediate feature map. During the forward pass, we do not store other feature maps, to save memory. Since the reconstructed map fits into GPU memory, it can subsequently be fed through the rest of the neural network, resulting in the final output. 

For the backward pass, we can use a similar implementation. The last layers, until the reconstructed feature map, can be backpropagated as usual. Then, we correctly tile the gradient of the feature map, with every gradient tile belonging to an input tile. Leveraging the input tile, we recalculate the features of the first layers with a partial forward pass (this is commonly called gradient checkpointing\cite{Chen2016a}). With the recalculated features and the gradient tile, we can finish the backpropagation of the network. This way, we can recover the gradients of all parameters, as would be the case if training with the original input image.

To train the ResNet-34, we streamed with a tile size of 2800$\times$2800 over the first 28 layers of the network. After these layers, the whole feature map (with dimensions 512$\times$512$\times$512) could fit into GPU memory. It is possible to use the streaming implementation for more layers of the network, however, to improve speed it is better to stream until the feature map is just small enough. Finally, we fed the map through the remaining six layers to calculate the final output.

\subsubsection{Gradient accumulation and parallelization}

Gradient accumulation is a technique to do a forward and backward pass on multiple images in series on the accelerator card, and averaging the parameter gradients over those images. Only after averaging, we perform a gradient descent step. Averaging the gradients over multiple images in series results in effectively training a mini-batch of these multiple images, while only requiring the memory for one image at a time. We used gradient accumulation over multiple biopsies to achieve an effective mini-batch size of 16 images. 

We trained over multiple GPUs by splitting the mini-batch. For the streaming experiments, we used four GPUs (either NVIDIA RTX 2080ti or GTX 1080ti).

\begin{figure*}[t]
    \centering
    \captionsetup{width=.8\textwidth,justification=centering}
    {\includegraphics[width=.8\textwidth]{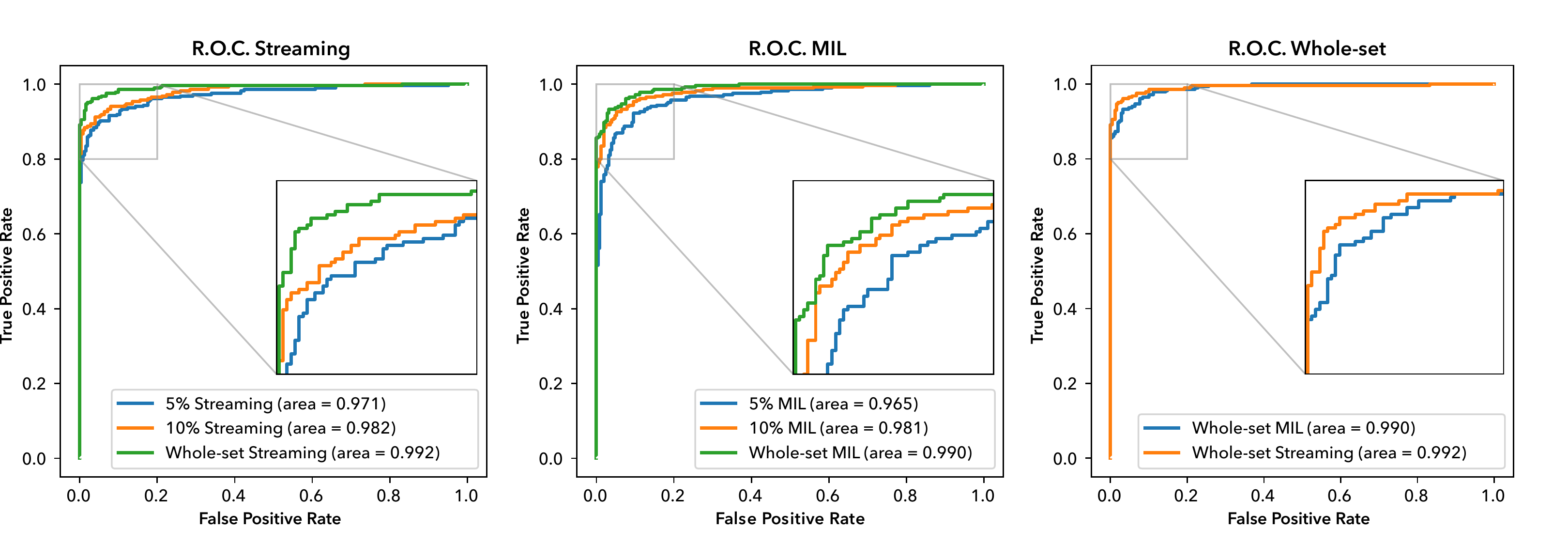}}%
    \caption{On the left, receiver-operating curves of streaming and MIL models, for all dataset sizes. On the right, comparison of receiver-operating curves of the two methods trained on the whole dataset.}
    \label{figure:ROCcomparison}
\end{figure*}

\subsection{Multiple-instance-learning model}

As a baseline, we implemented the multiple-instance-learning method as described in\cite{Campanella2019}. 

This method divides the images into a grid of smaller patches with the assumption that an individual patch could determine the image-level label. The task is to find the most informative patch. In our binary detection task, the most informative patch is determined by the patch with the highest probability of tumor. If there is a patch with a high probability of tumorous tissue, the whole biopsy is labeled tumorous.

We train such a model, per epoch, in two phases. The first phase is the inference phase, where we process all the patches of a biopsy, thereby finding the patch with the highest probability. This patch gets assigned the image-level label. Then, in the training phase, using only patches with the highest probability (the top-1 patch), the model parameters are optimized with a loss calculated on the patch probability and the label.

We followed the implementation from Campanella et al.\cite{Campanella2019}, but tweaked it for our dataset sizes. We used standard SGD (learning rate of $1e-5$) with momentum (0.9) with a mini-batch size of 16 images. We froze the BatchNormalization mean and variance, due to the smaller mini-batch size and to keep the features equal between the inference phase and the training phase. Equally, we oversampled negative cases to counter the imbalance in the dataset, instead of weighting\cite{Buda2018}.

We updated the whole model for 100 epochs. From these epochs, we picked the checkpoint with the lowest loss using the same scheme as the streaming model. Afterward, we trained for another 100 epochs with a learning rate of $3e-6$. We again choose the checkpoint based on the lowest tuning set loss, using a moving average of 5 epochs. We also used weight averaging for these checkpoints.

For regularization, we used the same data augmentation as the end-to-end model. We made sure that the same augmented patch was used in the inferencing and training phase. We used ImageNet statistics to normalize the patches.

\subsection{Quantitative evaluation}
The quantitative evaluation of both methods is performed using receiver-operating characteristic (ROC) analysis. Specifically, we look at the area under the ROC curve. To calculate a confidence interval for the test set, we used bootstrapping. We sampled 535 (size of the test set) biopsies, with replacement, from the test set, and calculated the area under the receiver-operating-curve based on the new sample. Repeating this procedure 10.000 times resulted in a distribution from which we calculated the 95\% confidence interval (2.5 and 97.5 percentile)

\subsection{Qualitative evaluation}

To assess the correlation of certain regions to the cancerous label, we created heatmaps for both techniques. For MIL, we used the patch probabilities. For streaming, we used sensitivity maps using SmoothGrad\cite{Smilkov2017}. As implementation of SmoothGrad, we averaged 25 sensitivity maps on Gaussian-noise-augmented versions of a biopsy. We used a standard deviation of 5\% of the image-wide standard deviation for the Gaussian noise. As a comparison, we show pixel-level segmentations from the model published in Bulten et al.\cite{Bulten2020} as well. 

In addition, we did a thorough analysis of the false positives and negatives of both the MIL and the streaming methods. 

\section{Experiments}

We performed three experiments for both methods using three datasets. One experiment on all the data, and two on subsampled training sets, the 10\% (500 biopsies) and 5\% (250 biopsies) datasets.

\begin{table}[h]
\footnotesize
\centering
\captionsetup{width=.85\linewidth,justification=centering}
\caption{Area under the receiver-operating-curve comparison between the methods on the test set.}
\label{tab:test_results}
\begin{tabular}{llll}
\toprule
\textbf{Dataset} & \textbf{Streaming} & \textbf{MIL} \\
\midrule
Whole set & 0.992 (0.985--0.997) & 0.990 (0.984--0.995) \\
\midrule
10\% set & 0.982 (0.972--0.990) & 0.981 (0.970--0.990) \\
5\% set & 0.971 (0.960--0.982) & 0.965 (0.949--0.978) \\
\midrule
\midrule
\textit{Bulten et al.\cite{Bulten2020}} & \textit{0.990 (0.982–-0.996)} & & \\
\bottomrule
\end{tabular}
\end{table}

On the whole dataset, the streaming model achieved an AUC of 0.992 (0.985–0.997) and the MIL model an AUC of 0.990 (0.984–0.995). Interestingly, our models trained on the whole dataset reached similar performance to previous work on this dataset\cite{Bulten2020}, which utilized a segmentation network trained using dense annotations obtained in a semi-supervised fashion.

For streaming, the performance on the smaller dataset sizes are similar between the two. 5\% dataset has an AUC of 0.971 for 5\% and 0.982 for 10\% (Table \ref{tab:test_results}). The models trained with more data generalize better (Fig. \ref{figure:ROCcomparison}).

Also for multiple-instance learning there is a clear improvement going from a model trained on the smallest dataset size, with an AUC of 0.965, increasing to 0.981 (0.970–0.990) on the 10\% dataset.

There seems to be a trend that the MIL model performs slightly worse (Fig. \ref{figure:ROCcomparison}), however, this difference falls within the confidence intervals. 

In general, the areas identified by MIL and streaming in the heatmaps correspond well to the pixel-level segmentations from Bulten et al., showing that both methods pick up the relevant regions for cancer identification (Figure \ref{figure:streamingSGD}). Most errors of the models seem to be due to normal epithelium mimicking tumorous glands in the case for false positives, and the small size of some tumorous regions as a possible reason for the false negatives. (Table \ref{tab:errors})

\begin{table}[h]
\footnotesize
\centering
\captionsetup{width=.9\linewidth,justification=centering}
\caption{Possible sources of errors for both models. The predictions were manually judged and divided in the following categories. False positives and negatives were selected at the point of maximum accuracy in the ROC curve.}
\label{tab:errors}
\begin{tabular}{lll}
\toprule
\textbf{False positives} & \textbf{Streaming} (5) & \textbf{MIL} (13) \\
\midrule
Normal mimicking tumor & 2 & 7 \\
Inflammation & 1 & 4 \\
Tissue artefacts & 1 & 1 \\
Bladder epithelium & 1 & 0 \\
Colon epithelium & 0 & 1 \\
\midrule
\textbf{False negatives} & \textbf{Streaming} (13) & \textbf{MIL} (12) \\
\midrule
Little amount of tumor & 7 & 4 \\
Tissue artefacts & 3 & 1 \\
Low-grade tumor & 1 & 2 \\
Inflammation-like & 1 & 2 \\
Unclear reason & 1 & 2 \\
\bottomrule
\end{tabular}
\end{table}

\begin{figure}
    \centering
    \subfloat[Identified by MIL model.]{\includegraphics[width=.4\columnwidth]{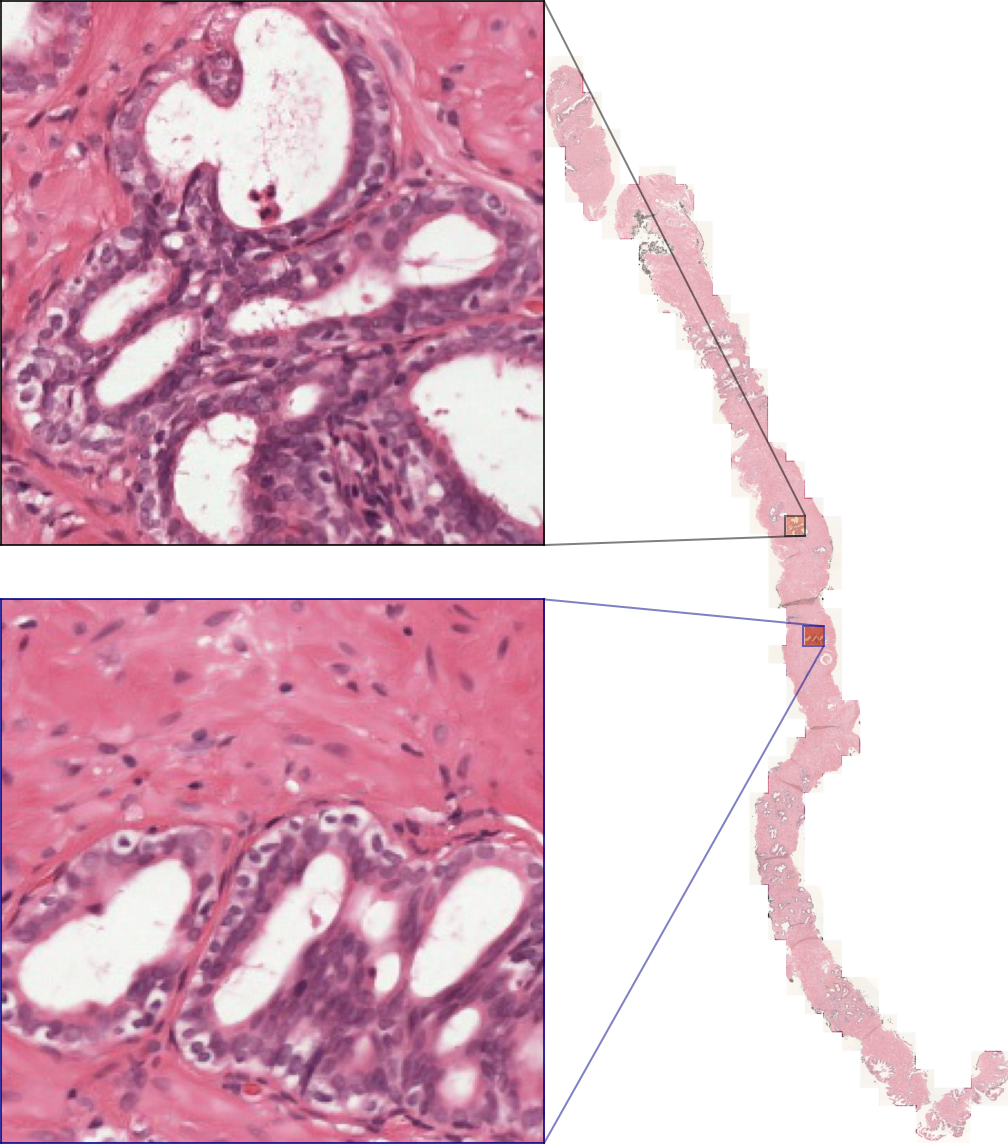}} %
    \qquad
    \subfloat[Identified by streaming model.]{\includegraphics[width=.37\columnwidth]{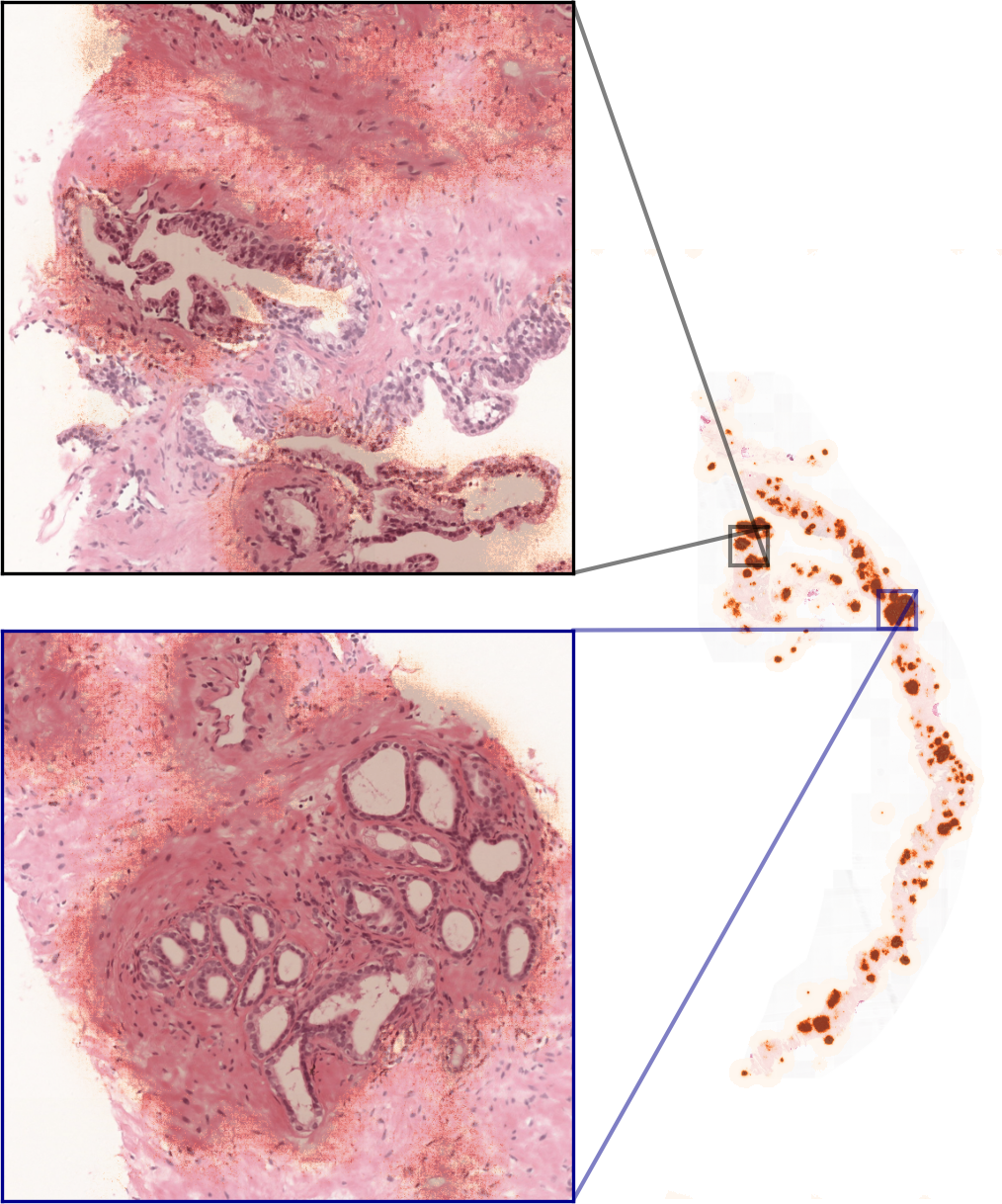}} %
    \caption{Examples of false positive predictions of test set biopsies, both small areas of normal epithelium that may resemble low-grade cancer. Showing patch probabilities for MIL (a), and SmoothGrad saliency for the streaming model (b), both overlayed on the overview biopsy. The zoomed-in region for MIL is exactly one patch.}%
    \label{fig:example}%
\end{figure}
\begin{figure}
    \centering
    \subfloat[Small tumorous glands mimicking vessels. Missed by both models.]{\includegraphics[width=.33\columnwidth]{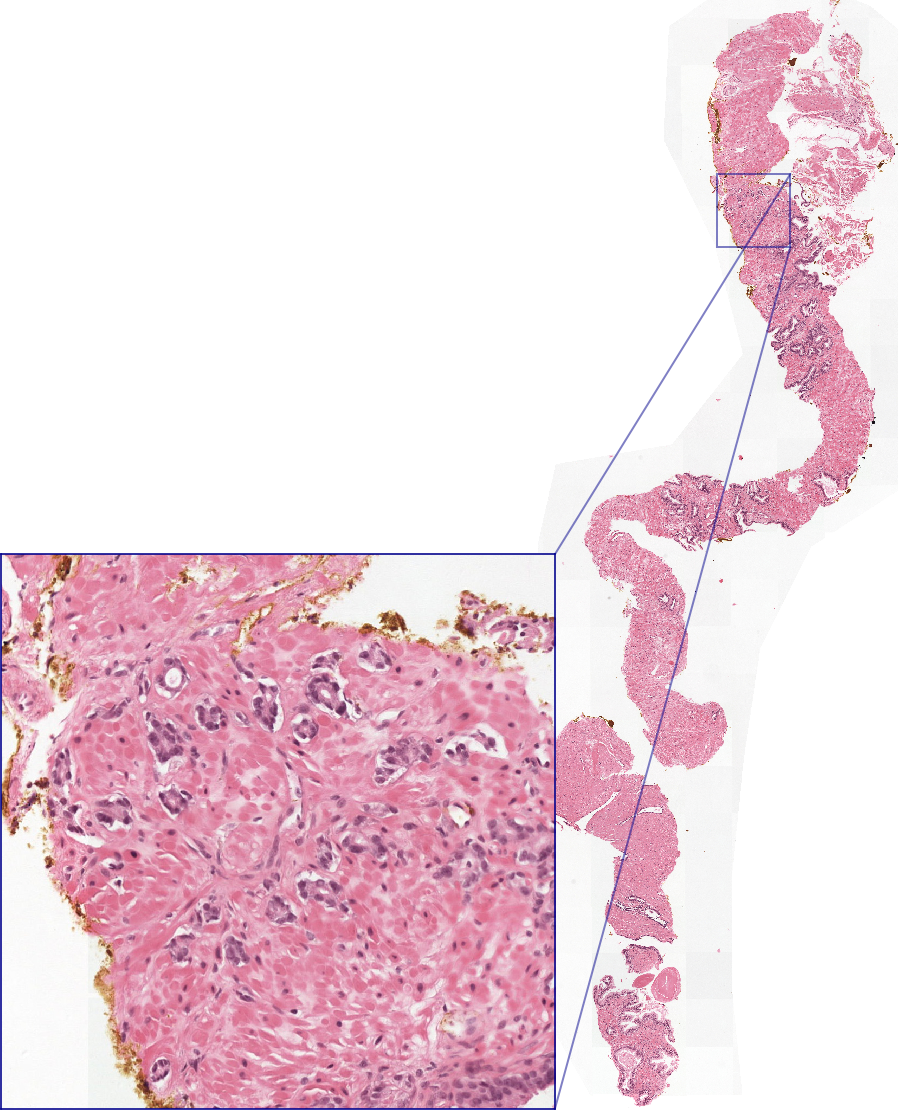}} %
    \qquad
    \qquad
    \subfloat[Very limited amount of tumor (four glands), missed by end-to-end network.]{\includegraphics[width=.37\columnwidth]{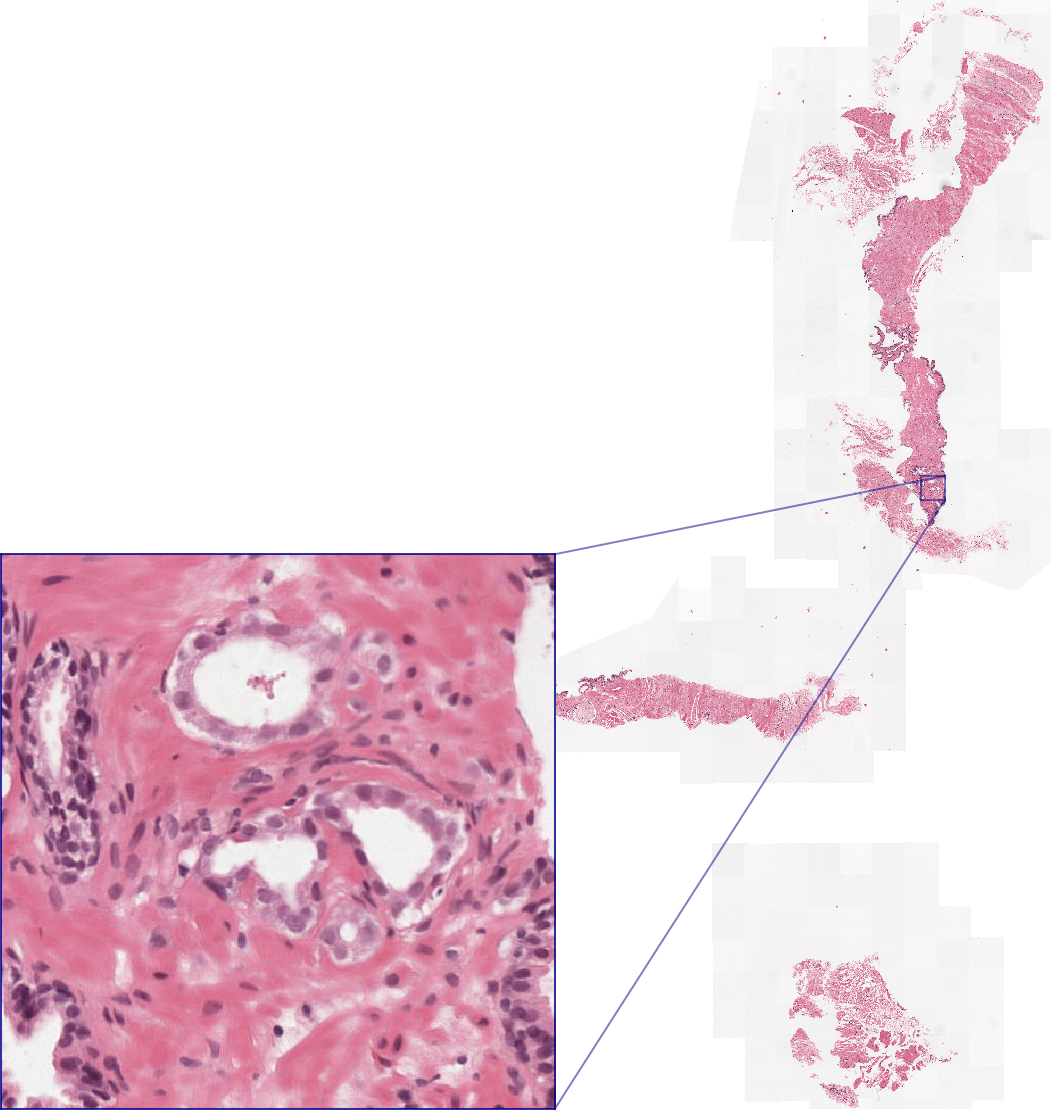}} %
    \caption{Examples of false negative predictions of test set biopsies with potential reasons for misclassification.}%
    \label{fig:example}%
\end{figure}

\begin{figure*} 
    \centering
    {\includegraphics[width=\textwidth]{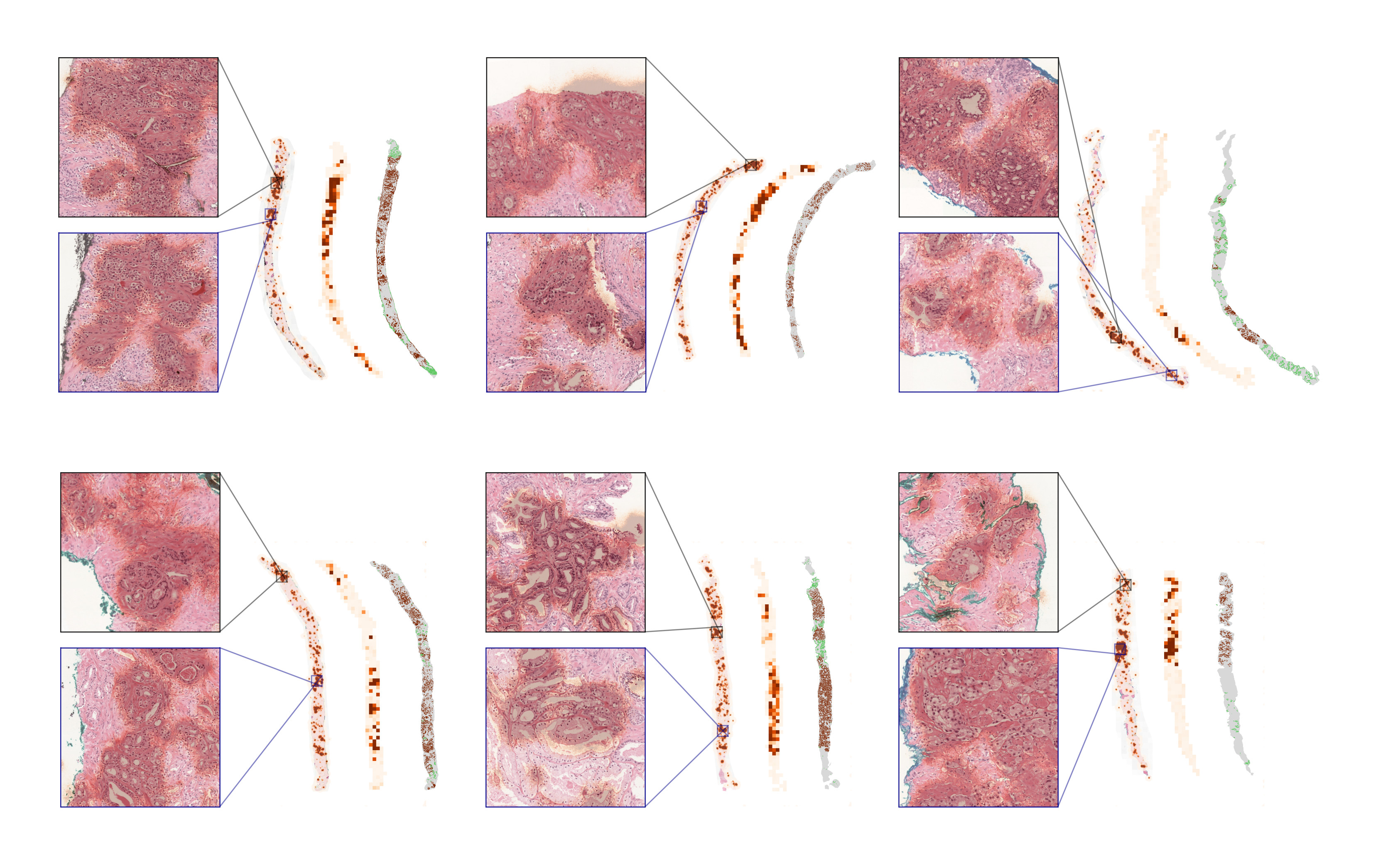}}%
    \caption{Heatmaps of the models trained on the whole dataset, for correctly predicted examples. The first biopsy shows a sensitivity map calculated using SmoothGrad. The second biopsy shows the probability per patch for the multiple-instance-learning model. The third biopsy shows a segmentation map from Bulten et al, 2020; healthy glands are denoted in green.}
    \label{figure:streamingSGD}
\end{figure*}

\section{Discussion and conclusions}

In this paper, we proposed using streaming\cite{Pinckaers2019} convolution neural networks to directly train a state-of-the-art ResNet\=/34 architecture on whole prostate biopsies with slide-level labels from pathology reports. We are the first to train such high-resolution (268 megapixels) images end-to-end, without further heuristics. Accomplishing this without the streaming implementation would require a accelerator card with 2.4 terabyte of memory.

We showed it is possible to train a residual neural network with biopsy level labels and reach similar performance to a popular multiple-instance-learning (MIL) based method. Our models trained on the whole dataset reached an AUC of 0.992 for streaming training, and 0.990 for MIL. In addition, we achieved equal performance to a method trained on patch-based labels, with an AUC of 0.990\cite{Bulten2020} on the same dataset. Although, it should be noted that Bulten et al. used weakly-supervised labels, they used a cascade of models to go from epithelium antibody-staining to semi-automatic pixel-level annotations, to generate a model trained at the patch level.

Looking at the failure cases (Table \ref{tab:errors}), multiple-instance-learning suffers from interpreting normal glands as tumorous. We hypothesize this is due to the lack of context, in all but one of these cases the misclassification was due to one patch. For false negatives, both models fail when there is a small amount of tumor, however the streaming model seems to suffer more from this. A possible solution would be to incorporate attention mechanisms into the network, allowing it to focus to smaller parts of the biopsy.

In this paper, we compared against a MIL implementation of Campanella et al. In their MIL implementation, only the top-1 patch is used for training per epoch. The method's data efficiency is reliant on how often different patches are selected in the first phase. Our results on the smallest dataset sample (5\%, 250 slides) hint towards reduced data efficiency for MIL. However, the performance on the smaller datasets was already close to optimal, suggesting effective use of the transferred ImageNet-weights. Even though it is not the same test set as in their original paper, this seems to suggest a better performance for smaller datasets than Campanella et al. reported. Hypothetically, this could be due to data augmentation, which they did not use, and increased randomness with smaller mini-batch size in our study.

For MIL, selecting different patches per image, every epoch, is important to circumvent overfitting. We used lower minibatch-sizes, 16 vs 512, and learning rates, $1e-5$ vs $1e-4$ as the original implementation\cite{Campanella2019}. We saw increased stability in training using smaller mini-batch sizes and learning rates, especially for the smaller datasets, where the whole dataset would otherwise fit in one mini-batch. Lower mini-batch sizes increased some noise, thereby picking different patches per epoch.

The streaming implementation of convolutional neural networks is computationally slower than our baseline. Mainly due to the number (121) and overlap (\textasciitilde 650 pixels) of the tiles during backpropagation. We improved training speed by first freezing the first layers of the neural network, not having to calculate gradients. Using this training scheme in the multiple-instance-learning baseline resulted in unstable training and worse performance. 

Streaming training with high-resolution images opens up the possibility to quickly gather large datasets with labels from pathology reports to train convolutional neural networks. Although linking individual biopsies to the pathology report is still a manual task, it is more efficient than annotating the individual slides. However, some pathology labs will manufacture one slide per biopsy and report systematically on these individual biopsies. Training from a whole slide, with multiple biopsies, is left for future research.

Since multiple-instance-learning, in the end, predicts the final label on a single patch, tasks that require information from different sites of the biopsy could be hard to engineer in this framework. For example, in the Gleason grading scheme, the two most informative growth patterns are reported. These patterns could lie on different parts of the biopsy, outside of the field-of-view of a single patch. Also, additional growth patterns could be present. The first reported growth pattern of Gleason grading is the most prevalent. Since multiple-instance-learning works patch-based, volumes that span larger than one patch are not used for the prediction. Streaming allows for training complex tasks, such as cancer grading, even with slide-level labels.

Our heatmaps show that indeed the streaming model uses information from multiple regions in the biopsy. Even though our model is not trained on a patch-level, the sensitivity maps highlight similar regions as the MIL method and the segmentation algorithm from Bulten et al. Thus, interestingly, a modern convolutional neural network, originally developed for tiny input sizes, can extract useful information from 268 megapixel images.

Besides allowing the entire slide to inform predictions, streaming training also has the advantage of being able to learn with hard or impossible to annotate global information. For example, in the medical domain, survival prediction can be of great interest. Future work could be to predict survival from histopathology tissue directly. Reliably annotating for this task can be difficult. Since streaming can find patterns and features from the whole image using just the retrospective patient prognosis, this method can be beneficial in automatically finding new relevant biomarkers.

We provide source code of the streaming pipeline at GitHub\footnote{https://github.com/DIAGNijmegen/pathology-streaming-pipeline}. We tried to make it easy to use with other datasets. Additionally to methods used in this paper, we added mixed precision support for even more memory efficient and faster training.

\bibliographystyle{ieeetr}
\bibliography{library}

\begin{thebibliography}{10}

\bibitem{Krizhevsky2009}
A.~Krizhevsky, ``{Learning Multiple Layers of Features from Tiny Images},''
  tech. rep., University of Toronto, 2009.

\bibitem{Russakovsky2015}
O.~Russakovsky, J.~Deng, H.~Su, J.~Krause, S.~Satheesh, S.~Ma, Z.~Huang,
  A.~Karpathy, A.~Khosla, M.~Bernstein, A.~C. Berg, and L.~Fei-Fei, ``{ImageNet
  Large Scale Visual Recognition Challenge},'' {\em International Journal of
  Computer Vision}, vol.~115, pp.~211--252, dec 2015.

\bibitem{Sun2017RevisitingUE}
C.~Sun, A.~Shrivastava, S.~Singh, and A.~Gupta, ``{Revisiting Unreasonable
  Effectiveness of Data in Deep Learning Era},'' {\em 2017 IEEE International
  Conference on Computer Vision (ICCV)}, pp.~843--852, 2017.

\bibitem{Fine2012}
S.~W. Fine, M.~B. Amin, D.~M. Berney, A.~Bjartell, L.~Egevad, J.~I. Epstein,
  P.~A. Humphrey, C.~Magi-Galluzzi, R.~Montironi, and C.~Stief, ``{A
  contemporary update on pathology reporting for prostate cancer: Biopsy and
  radical prostatectomy specimens},'' {\em European Urology}, vol.~62, no.~1,
  pp.~20--39, 2012.

\bibitem{Ozkan2016}
T.~A. Ozkan, A.~T. Eruyar, O.~O. Cebeci, O.~Memik, L.~Ozcan, and I.~Kuskonmaz,
  ``{Interobserver variability in Gleason histological grading of prostate
  cancer},'' {\em Scandinavian Journal of Urology}, vol.~50, pp.~420--424, nov
  2016.

\bibitem{10.1093/jnci/djp278}
H.~G. Welch and P.~C. Albertsen, ``{Prostate Cancer Diagnosis and Treatment
  After the Introduction of Prostate-Specific Antigen Screening:
  1986–2005},'' {\em JNCI: Journal of the National Cancer Institute},
  vol.~101, no.~19, pp.~1325--1329, 2009.

\bibitem{Wilson2018}
M.~L. Wilson, K.~A. Fleming, M.~A. Kuti, L.~M. Looi, N.~Lago, and K.~Ru,
  ``{Access to pathology and laboratory medicine services: a crucial gap.},''
  {\em Lancet (London, England)}, vol.~391, pp.~1927--1938, may 2018.

\bibitem{Bulten2020}
W.~Bulten, H.~Pinckaers, H.~van Boven, R.~Vink, T.~de~Bel, B.~van Ginneken,
  J.~van~der Laak, C.~{Hulsbergen-van de Kaa}, and G.~Litjens, ``{Automated
  deep-learning system for Gleason grading of prostate cancer using biopsies: a
  diagnostic study},'' {\em The Lancet Oncology}, vol.~21, pp.~233--241, feb
  2020.

\bibitem{Campanella2019}
G.~Campanella, M.~G. Hanna, L.~Geneslaw, A.~Miraflor, V.~{Werneck Krauss
  Silva}, K.~J. Busam, E.~Brogi, V.~E. Reuter, D.~S. Klimstra, and T.~J. Fuchs,
  ``{Clinical-grade computational pathology using weakly supervised deep
  learning on whole slide images},'' {\em Nature Medicine}, vol.~25,
  pp.~1301--1309, aug 2019.

\bibitem{Nagpal2019}
K.~Nagpal, D.~Foote, Y.~Liu, P.-H.~C. Chen, E.~Wulczyn, F.~Tan, N.~Olson, J.~L.
  Smith, A.~Mohtashamian, J.~H. Wren, G.~S. Corrado, R.~MacDonald, L.~H. Peng,
  M.~B. Amin, A.~J. Evans, A.~R. Sangoi, C.~H. Mermel, J.~D. Hipp, and M.~C.
  Stumpe, ``{Development and validation of a deep learning algorithm for
  improving Gleason scoring of prostate cancer},'' {\em npj Digital Medicine},
  vol.~2, p.~48, dec 2019.

\bibitem{Litjens2016}
G.~Litjens, C.~I. S{\'{a}}nchez, N.~Timofeeva, M.~Hermsen, I.~Nagtegaal,
  I.~Kovacs, C.~{Hulsbergen - van de Kaa}, P.~Bult, B.~van Ginneken, and
  J.~van~der Laak, ``{Deep learning as a tool for increased accuracy and
  efficiency of histopathological diagnosis},'' {\em Scientific Reports},
  vol.~6, p.~26286, sep 2016.

\bibitem{Arvaniti2018}
E.~Arvaniti, K.~S. Fricker, M.~Moret, N.~Rupp, T.~Hermanns, C.~Fankhauser,
  N.~Wey, P.~J. Wild, J.~H. R{\"{u}}schoff, and M.~Claassen, ``{Automated
  Gleason grading of prostate cancer tissue microarrays via deep learning},''
  {\em Scientific Reports}, vol.~8, p.~12054, dec 2018.

\bibitem{Lucas2019}
M.~Lucas, I.~Jansen, C.~D. Savci-Heijink, S.~L. Meijer, O.~J. de~Boer, T.~G.
  van Leeuwen, D.~M. de~Bruin, and H.~A. Marquering, ``{Deep learning for
  automatic Gleason pattern classification for grade group determination of
  prostate biopsies},'' {\em Virchows Archiv}, vol.~475, pp.~77--83, jul 2019.

\bibitem{Strom2020}
P.~Str{\"{o}}m, K.~Kartasalo, H.~Olsson, L.~Solorzano, B.~Delahunt, D.~M.
  Berney, D.~G. Bostwick, A.~J. Evans, D.~J. Grignon, P.~A. Humphrey, K.~A.
  Iczkowski, J.~G. Kench, G.~Kristiansen, T.~H. van~der Kwast, K.~R.~M. Leite,
  J.~K. McKenney, J.~Oxley, C.-C. Pan, H.~Samaratunga, J.~R. Srigley,
  H.~Takahashi, T.~Tsuzuki, M.~Varma, M.~Zhou, J.~Lindberg, C.~Lindskog,
  P.~Ruusuvuori, C.~W{\"{a}}hlby, H.~Gr{\"{o}}nberg, M.~Rantalainen, L.~Egevad,
  and M.~Eklund, ``{Artificial intelligence for diagnosis and grading of
  prostate cancer in biopsies: a population-based, diagnostic study},'' {\em
  The Lancet Oncology}, vol.~21, pp.~222--232, feb 2020.

\bibitem{Courtiol2018}
P.~Courtiol, E.~W. Tramel, M.~Sanselme, and G.~Wainrib, ``{Classification and
  Disease Localization in Histopathology Using Only Global Labels: A
  Weakly-Supervised Approach},'' {\em arXiv preprint}, vol.~1802.02212, 2018.

\bibitem{Ilse2018}
M.~Ilse, J.~M. Tomczak, and M.~Welling, ``{Attention-based Deep Multiple
  Instance Learning},'' {\em arXiv preprint}, vol.~1802.04712, 2018.

\bibitem{Amores2013}
J.~Amores, ``{Multiple instance classification: Review, taxonomy and
  comparative study},'' {\em Artificial Intelligence}, vol.~201, pp.~81--105,
  aug 2013.

\bibitem{VanderLaak2019}
J.~van~der Laak, F.~Ciompi, and G.~Litjens, ``{No pixel-level annotations
  needed},'' {\em Nature Biomedical Engineering}, vol.~3, pp.~855--856, oct
  2019.

\bibitem{Pinckaers2019}
H.~Pinckaers, B.~van Ginneken, and G.~Litjens, ``{Streaming convolutional
  neural networks for end-to-end learning with multi-megapixel images},'' {\em
  arXiv preprint}, vol.~1911.04432, 2019.

\bibitem{Gertych2015}
A.~Gertych, N.~Ing, Z.~Ma, T.~J. Fuchs, S.~Salman, S.~Mohanty, S.~Bhele,
  A.~Vel{\'{a}}squez-Vacca, M.~B. Amin, and B.~S. Knudsen, ``{Machine learning
  approaches to analyze histological images of tissues from radical
  prostatectomies},'' {\em Computerized Medical Imaging and Graphics}, vol.~46,
  pp.~197--208, 2015.

\bibitem{nguyen2017}
T.~H. Nguyen, S.~Sridharan, V.~Macias, A.~Kajdacsy-Balla, J.~Melamed, M.~N. Do,
  and G.~Popescu, ``{Automatic Gleason grading of prostate cancer using
  quantitative phase imaging and machine learning},'' {\em Journal of
  biomedical optics}, vol.~22, no.~3, p.~36015, 2017.

\bibitem{Naik2007}
S.~Naik, S.~Doyle, M.~Feldman, J.~Tomaszewski, and A.~Madabhushi, ``{Gland
  Segmentation and Computerized Gleason Grading of Prostate Histology by
  Integrating Low- , High-level and Domain Specific Information.},'' in {\em
  Proceedings of 2nd Workshop on Microsopic Image Analysis with Applications in
  Biology}, pp.~1--8, 2007.

\bibitem{Ianni2020}
J.~D. Ianni, R.~E. Soans, S.~Sankarapandian, R.~V. Chamarthi, D.~Ayyagari,
  T.~G. Olsen, M.~J. Bonham, C.~C. Stavish, K.~Motaparthi, C.~J. Cockerell,
  T.~A. Feeser, and J.~B. Lee, ``{Tailored for Real-World: A Whole Slide Image
  Classification System Validated on Uncurated Multi-Site Data Emulating the
  Prospective Pathology Workload},'' {\em Scientific Reports}, vol.~10, no.~1,
  p.~3217, 2020.

\bibitem{lu2020data}
M.~Y. Lu, D.~F.~K. Williamson, T.~Y. Chen, R.~J. Chen, M.~Barbieri, and
  F.~Mahmood, ``{Data Efficient and Weakly Supervised Computational Pathology
  on Whole Slide Images},'' 2020.

\bibitem{Li2019a}
J.~Li, W.~Li, A.~Gertych, B.~S. Knudsen, W.~Speier, and C.~W. Arnold, ``{An
  attention-based multi-resolution model for prostate whole slide
  imageclassification and localization},'' {\em arXiv preprint},
  vol.~1905.13208, 2019.

\bibitem{Tellez2019}
D.~Tellez, G.~Litjens, J.~van~der Laak, and F.~Ciompi, ``{Neural Image
  Compression for Gigapixel Histopathology Image Analysis},'' {\em IEEE
  Transactions on Pattern Analysis and Machine Intelligence}, vol.~in press.

\bibitem{Bandi2019a}
P.~B{\'{a}}ndi, M.~Balkenhol, B.~van Ginneken, J.~van~der Laak, and G.~Litjens,
  ``{Resolution-agnostic tissue segmentation in whole-slide histopathology
  images with convolutional neural networks.},'' {\em PeerJ}, vol.~7, p.~e8242,
  2019.

\bibitem{He2016}
K.~He, X.~Zhang, S.~Ren, and J.~Sun, ``{Deep Residual Learning for Image
  Recognition},'' in {\em 2016 IEEE Conference on Computer Vision and Pattern
  Recognition (CVPR)}, pp.~770--778, 2016.

\bibitem{VIPS1996}
J.~Cupitt and K.~Martinez, ``{VIPS: An imaging processing system for large
  images},'' {\em Proceedings of SPIE - The International Society for Optical
  Engineering}, vol.~1663, pp.~19 -- 28, 1996.

\bibitem{Buda2018}
M.~Buda, A.~Maki, and M.~A. Mazurowski, ``{A systematic study of the class
  imbalance problem in convolutional neural networks.},'' {\em Neural networks
  : the official journal of the International Neural Network Society},
  vol.~106, pp.~249--259, oct 2018.

\bibitem{TanSKZYL18}
C.~Tan, F.~Sun, T.~Kong, W.~Zhang, C.~Yang, and C.~Liu, ``{A Survey on Deep
  Transfer Learning},'' in {\em Artificial Neural Networks and Machine Learning
  - {\{}ICANN{\}} 2018 - 27th International Conference on Artificial Neural
  Networks, Rhodes, Greece, October 4-7, 2018, Proceedings, Part {\{}III{\}}},
  vol.~11141 of {\em Lecture Notes in Computer Science}, pp.~270--279,
  Springer, 2018.

\bibitem{Izmailov2018}
P.~Izmailov, D.~Podoprikhin, T.~Garipov, D.~Vetrov, and A.~G. Wilson,
  ``{Averaging Weights Leads to Wider Optima and Better Generalization},'' {\em
  arXiv preprint}, vol.~1803.05407, 2018.

\bibitem{Chen2016a}
T.~Chen, B.~Xu, C.~Zhang, and C.~Guestrin, ``{Training Deep Nets with Sublinear
  Memory Cost},'' {\em arXiv preprint}, vol.~1604.06174, apr 2016.

\bibitem{Smilkov2017}
D.~Smilkov, N.~Thorat, B.~Kim, F.~B. Vi{\'{e}}gas, and M.~Wattenberg,
  ``{SmoothGrad: removing noise by adding noise},'' in {\em ICML workshop on
  Visualization for deep learning}, 2017.

\end{thebibliography}

\end{document}